\documentclass[fleqn,usenatbib]{mnras}
\usepackage{newtxtext,newtxmath}

\usepackage[T1]{fontenc}
\DeclareRobustCommand{\VAN}[3]{#2}
\let\VANthebibliography\thebibliography
\def\thebibliography{\DeclareRobustCommand{\VAN}[3]{##3}\VANthebibliography}

\usepackage{graphicx}	
\usepackage{amsmath}	
\usepackage{subcaption}
\usepackage{xspace}

\newif\iffirstRS
\firstRStrue 

\newcommand{\RS}{%
  \iffirstRS
    2000~RS$_{11}$%
    \firstRSfalse
  \else
    RS$_{11}$%
  \fi
}

\newcommand{\DP}{DP$_{14}$}
\newcommand{\JV}{JV$_{6}$}

\newcommand{\HW}{HW$_{1}$}
\newcommand{\KZ}{KZ$_{66}$}

\newcommand{\shape}{\texttt{SHAPE}}

\title[The shape of (275677) 2000 \RS]{The shape and spin state of (275677) \RS~from ground-based radar and optical observations}

\author[R. E. Cannon et al.]{Richard E. Cannon$^{1}$\thanks{{\bf E-mail:} richard.cannon@ed.ac.uk {\bf \& ORCID:} 0009-0007-5946-8731},
Agata Ro\.{z}ek$^{1}$, 
Kaley Brauer$^{2}$, 
Michael W. Busch$^{3}$, 
Colin Snodgrass$^{1}$, \newauthor 
Lance A. M. Benner$^{4}$, 
Marina Brozovi\'{c}$^{4}$, 
Jon D. Giorgini$^{4}$, 
Ellen Howell$^{5}$, 
Michael C. Nolan$^{5}$,\newauthor 
Markus Rabus$^{6}$, 
Sedighe Sajadian$^{7}$, 
Alessondra Springmann$^{8}$, 
Patrick A. Taylor$^{9}$,\newauthor 
Luisa Fernanda Zambrano-Marin$^{10}$ 
\\
$^{1}$Institute for Astronomy, University of Edinburgh, Royal Observatory, Edinburgh, EH9 3HJ, UK \\
${^2}$Center for Astrophysics, Harvard University, 60 Garden St, Cambridge, MA 02138, USA\\
$^{3}$SETI Institute, Mountain View, CA 94043, USA \\
$^{4}$Jet Propulsion Laboratory, California Institute of Technology, Pasadena, CA 91109, USA \\
$^{5}$Lunar and Planetary Laboratory, University of Arizona, Tucson, AZ 85721, USA\\
$^{6}$Departamento de Matemática y Física Aplicadas, Facultad de Ingeniería, Universidad Católica de la Santísima Concepción, Alonso de Rivera 2850,\\ Casilla 297 Concepción, Chile\\
$^{7}$Perimeter Institute for Theoretical Physics, Waterloo, ON N2L 2Y5, Canada\\
$^{8}$Sonoran Space Research, Boulder, CO 80304, USA\\
$^{9}$National Radio Astronomy Observatory, Charlottesville, VA 22903, USA\\
$^{10}$Florida Space Institute, University of Central Florida, 12354 Research Pkwy., Orlando, FL 32826, USA\\
}

\date{Accepted XXX. Received YYY; in original form ZZZ}

\pubyear{\the\year{}}

\begin{document}
\label{firstpage}
\pagerange{\pageref{firstpage}--\pageref{lastpage}}
\maketitle

\begin{abstract}
Near-Earth asteroid (275677) \RS~was observed over 5 days in March 2014 with both the Arecibo ($2380$ MHz, $12.6$ cm) and Goldstone ($8560$ MHz, $3.5$ cm) planetary radar systems. 
The continuous-wave spectra and delay-Doppler images collected revealed a sub-km-sized object with a strongly bifurcated shape. 
We used these radar observations, in combination with $7$ optical lightcurves collected in 2014 and one lightcurve from 2023, to create a comprehensive shape and spin-state model for \RS.   
We find a rotation period of $P = (4.445\pm0.001)$ hours around a pole of $\lambda = (225\pm80)^{\circ}$ and $\beta = (-80\pm9)^{\circ}$ relative to the plane of the ecliptic.
The shape of \RS~is unusual in that it does not resemble many of the other near-Earth asteroids modelled with ground-based radar.
Whilst \RS~consists of a largely spherical, smaller lobe attached to an elongated, larger lobe via a narrow neck, the smaller lobe is not aligned with the long axis of the larger lobe, but is closer to the larger lobe's shortest principal axis.
In combination with a large concavity observed on the outer face of the larger lobe, this may point to an unusual formation or event in the object's past.
We estimate that \RS~has an geometric albedo of $p_v = (0.16 \pm 0.06)$ and a radar albedo $0.08 < \eta_{\rm OC} < 0.16$.
Analysis of its gravitational environment reveals that for standard S-type asteroid densities, we would not expect rotational instability and it is possible for \RS~to be a low tensile strength rubble-pile asteroid.

\end{abstract}




\section{Introduction} \label{sec:Introduction}

Near-Earth object (275677) \RS, henceforth \RS, is an Apollo group asteroid with an eccentricity of $0.32$ and a semi-major axis of $1.28$ au, crossing both Earth and Mars' orbits.
Classified as an Sa-type \citep{LazzarinEtAl2004_RS11-spectral-type} potentially hazardous asteroid due to its close approaches with Earth, it was discovered by LINEAR in the year 2000 and observed further in 2014, when it came within $0.05$ au ($< 20$ times the distance to the Moon) of Earth.
This 2014 approach enabled the collection of not only high-quality lightcurves but also high-quality radar observations from both the Goldstone and Arecibo observatories.

Observations of \RS~in 2014 measured a rotation period of $(4.444\pm0.001)$ h \citep{Warner2014_Lightcurves_dp14_rs11, Benishek2014_Lightcurves_rs11, Karavaev2015_RS11-lightcurve-modelling} with a large amplitude of up to $1.31$ mag at specific viewing geometries. 
Using their lightcurves, \citet{Karavaev2015_RS11-lightcurve-modelling} was able to approximate the data by modelling \RS~as an elongated spheroid with two large concavities.
Meanwhile, a visual inspection of the radar images collected the week prior strongly indicated an irregular bilobate morphology, with possible additional concavities on one lobe.
\citet{BrauerEtAl2014_ConfAbs2000rs11_shape} used the radar delay-Doppler images to start the process of creating a shape model of \RS, finding that it resembled a `rubber duck' shape similar to that of 67P/Churyumov–Gerasimenko \citep{SierksEtAl2015_67P-Shape}.

Such bilobed shapes are referred to as contact binaries, with radar imaging estimating that they comprise 15-30\% of near-Earth objects (NEOs) \citep{BennerEtAl2015A4_15-percent-CB-NEO, VirkkiEtAl2022_30percent-CB}.
The first detailed report of a contact binary was (4769) Castalia by \citet{OstroEtAl1990_Castalia_radar_imaging} and \citet{Hudson&Ostro1994_Castalia-Shape}.
Most confirmed contact binaries in the NEO population are models created from radar, such as (8567) 1996 \HW, (388188) 2006 \DP, or (85990) 1999 \JV~\citep{MagriEtAl2011_1996HW1-shape-CB, CannonEtAl2025_DP14shape, RozekEtAl2019_JV6-shape}, but contact binaries can also be clearly seen in radar observations that have not been modelled yet \citep[e.g.][]{BennerEtAl2006_2005cr37_radar_imaging, VirkkiEtAl2022_30percent-CB}.
The only contact binaries that have had radar models confirmed with spacecraft visits are (25143) Itokawa and (4179) Toutatis, which, whilst initially modelled with radar \citep{OstroEtAl2005_Itokawa_Radar_Model, HudsonEtAl2003_Toutatis-shape-CB}, were later visited by the Hayabusa and Chang'e-2 spacecraft respectively and imaged directly to produce more accurate shape models \citep{DemuraEtAl2006_Itokawa-Shape, ZhuEtAl2014_Toutatis-ChangE2-Shape, ZhaoEtAl2016_Toutatis-Shape-Radar-Spacecraft}.

Contact binaries do not exist solely in the NEO population.
Surveys of lightcurves in the main belt find the presence of high amplitude lightcurves that can signify elongated or contact binary shapes \citep{MasieroEtAl2009_MBA-1000-survey}.
Before the NASA Lucy mission, the only confirmed contact binary in the main belt was Kleopatra, which is large enough to have been both directly imaged and observed with radar observations from Earth \citep{OstroEtAl2000_KleopatraInitialRadarModel, ShepardEtAl2018_KleopatraShapeModel}.
In 2024, Lucy performed a flyby of the main belt asteroid (152830) Dinkinesh and discovered the first ever known case of a contact binary orbiting another asteroid, in the form of Dinkinesh's moon Selam \citep{LevisonEtAl2024_LUCY-Dinkinesh-Characterisation-Selam-first-look}, which is similar in size to known NEO contact binaries.
On the 20th of April in 2025, Lucy performed its second flyby of asteroid (52246) Donaldjohanson, which also appears to have a contact binary structure \citep{ScullyEtAl2025_EPSC-DJ-shape}.

Travelling further away from the Sun, extensive surveys of trans-Neptunian object (TNO) lightcurves suggest that 40-50\% of plutinos could be strongly elongated or bilobate \citep{Thirouin&Sheppard2018_40-percent-CB-in-Plutinos, PorterEtAl2020_ConfAbs-TNO-shapes-from-NH, Brunini2023_50-percent-CB-Plutinos-reason}, with similar or slightly lower proportions in other TNO families \citep{ThirouinSheppard2019_KBO-survey_10-25-pct-CB}.
Though high, these numbers may still be underestimates, due to the measured lightcurve magnitude's reliance on pole direction \citep{Lacerda2011_QG298-lc-change_CB-obliquity-correction, ShowalterEtAl2021_CBs-vs-Ellipsoids-TNO}.
The most notable TNO in this regard is (486958) Arrokoth, which was visited by the New Horizons mission \citep{KeaneEtAl2022_Arrokoth-Shape-Flyby}, but there are numerous other confirmed or suspected contact binaries greater than 25 km in size \citep{Sheppard&Jewitt2004_2001QG298-CB-KBO, ThirouinEtAl2017_2004tt357-CB-TNO, ThirouinSheppard2017_2002cc249-CB-TNO-KBO}.

Finally, contact binaries are found in comet populations. 
The nucleus of 67P, visited by the ESA Rosetta mission in 2014, has a distinctive bilobate shape with a defined neck feature \citep{SierksEtAl2015_67P-Shape}.
Other contact binary comet nuclei observed with spacecraft are 19P/Borrelly, visited by NASA's Deep Space 1 mission \citep{OberstEtAl2004_19P-spacecraft-images-shape}, and 103P/Hartley 2, visited by the EPOXI mission \citep{ThomasEtAl_103P-spacecraft-images-shape}.
These comet nuclei are similar in that they both have more elongated `bowling pin' shapes with indistinctive necks, unlike 67P.
On the other hand, comet 8P/Tuttle is one of the few comets that has been observed with ground-based radar, approaching close enough to Earth in 2008, and appears to have a narrower neck with respect to the width of its lobes \citep{HarmonEtAl2010_8P_radar_paper}.

Despite their prevalence, there are only 19 published contact binary shape models, most from either spacecraft missions or ground-based radar observations.
A complete list of these objects, with the recent exception of Donaldjohanson, can be found in \citet{CannonEtAl2025_DP14shape}.
The small number of models and the wide variety of their published shapes highlight a need for additional models to better understand the shapes and sizes of contact binary objects. 
In this paper, we present a new shape model of contact binary \RS~created using a selection of the lightcurves collected in 2014, an additional short lightcurve collected in 2023, and the radar delay-Doppler and continuous wave data. 
We first describe the observations used in Section~\ref{sec:Observations}, before discussing the modelling process in Section~\ref{sec:Modelling}. 
In Section~\ref {sec:Discussion}, we discuss the new model in relation to both the prior model and other contact binary models, and the implications for \RS's gravitational environment and rotational stability. 
Finally, we discuss the implications we can draw from this model regarding its possible formation and evolutionary history.

\section{Observations} \label{sec:Observations}

\begin{table*}
    \centering
    \caption{A list of all \RS~optical lightcurve observations. 
    Provided are the UT dates for the evening of the beginning of observations; 
    the site codes; 
    the length of each lightcurve, in hours; 
    the distance from the observer to the target, $\Delta$; 
    the solar phase angle, $\alpha$; 
    the observer centred ecliptic longitude, $\lambda$, and latitude, $\beta$; 
    and the filter used (IDs 1-7 were taken unfiltered, and then adjusted to be in Johnson V).
    The site MPC codes are U82 -- Palmer Divide Station, USA; K90 -- Sopot Astronomical Observatory, Serbia; W74 -- 1.54m Danish Telescope, La Silla, Chile. 
    Lightcurve IDs 1-7 were taken from ALCDEF \citep{WarnerEtAl2011_ALCDEFconcept}. 
    All lightcurves were used at all stages of the modelling process.}
    \label{tab:AllLightcurves}
    \begin{tabular}{cccccccccccc} \hline \hline
        ID & Date & Site & Length & $\Delta$ & $\alpha$ & $\lambda$ & $\beta$ & Filt. & Source & \\ 
        ~  & ~ & ~ & [h] & [au] & [\textdegree] & [\textdegree] & [\textdegree] & ~ & ~ & \\ \hline
		1 & 2014-03-16 & U82 & 4.1 & 0.048 & 73.5 & 248.3 & 37.4 & V & \citet{Warner2014_Lightcurves_dp14_rs11} \\
		2 & 2014-03-17 & U82 & 4.4 & 0.053 & 69.9 & 243.6 & 41.1 & V & \citet{Warner2014_Lightcurves_dp14_rs11} \\
		3 & 2014-03-19 & K90 & 2.9 & 0.061 & 65.1 & 236.6 & 45.5 & V & \citet{Benishek2014_Lightcurves_rs11} \\
		4 & 2014-03-20 & K90 & 2.0 & 0.067 & 62.8 & 232.8 & 47.5 & V & \citet{Benishek2014_Lightcurves_rs11} \\
		5 & 2014-03-21 & K90 & 5.1 & 0.072 & 61.0 & 229.6 & 49.0 & V & \citet{Benishek2014_Lightcurves_rs11} \\
		6 & 2014-03-22 & K90 & 3.1 & 0.078 & 59.2 & 226.3 & 50.3 & V & \citet{Benishek2014_Lightcurves_rs11} \\
		7 & 2014-03-23 & K90 & 2.1 & 0.084 & 57.7 & 223.4 & 51.3 & V & \citet{Benishek2014_Lightcurves_rs11} \\
		8 & 2023-07-12 & W74 & 1.1 & 0.777 & 23.0 & 319.2 & 30.0 & R & This work \\ \hline
    \end{tabular}
\end{table*}

\begin{table*}
    \centering
    \caption{A list of all radar observations used in shape modelling. We collected observations with both the DSN Goldstone DSS-14 antenna (label: G, observing at $8560$ MHz) and the Arecibo radar antenna (label: A, observing at $2380$ MHz). We did not adjust the displayed resolutions for differences in transmitter frequencies. Columns show the date and time of observations (UTC), the type of observation (CW -- continuous wave and DD -- delay-Doppler), the baud length and samples per baud (spb) of observations, the resolution per pixel of both delay and Doppler axes, the number of runs per observation (each run is a single transmit and receive cycle), the mean distance to the target, the mean sub-radar latitude (derived from the final pole solution), and the orbital solution used (Sol.).}
    \label{tab:RadarInfo}
    \begin{tabular}{cccccccccccccc} \hline \hline
        Tel. & Date       & Start-Stop          & Type  & Baud     & Spb & \multicolumn{2}{c}{Resolution} & Runs   & Distance & Sub-radar             & Sol. \\
        ~    & ~          & [hh:mm:ss-hh:mm:ss] & ~     & [$\mu$s] &     & [m]  & [Hz]                    & ~      & [au]     & latitude [$^{\circ}$] & ~    \\ \hline
        
		  G    & 2014-03-12 & 10:50:41-11:02:23   & CW    & --       & --  & --   & 0.5                     & 10     & 0.036    & 4.6                   & 58   \\
          
        G    & 2014-03-13 & 11:10:05-11:17:13   & CW    & --       & --  & --   & 0.5                     & 6      & 0.038    & 12.1                  & 62   \\
		  ~    & ~          & 11:33:40-12:08:08   & DD    & 0.25     & 1   & 37.5 & 0.48                    & 27     & 0.038    & 12.4       & ~ \\
        ~    & ~          & 12:54:03-13:15:29   & DD    & 0.25     & 1   & 37.5 & 0.48                    & 14     & 0.038    & 12.7       & ~ \\
        
        A    & 2014-03-14 & 09:34:35-10:03:00   & CW    & --       & --  & --   & 0.0303                  & 20     & 0.040    & 18.3       & ~ \\
		  ~    & ~          & 10:14:00-11:12:57   & DD    & 0.05     & 1   & 7.5  & 0.074                   & 35     & 0.040    & 18.5                  & 66   \\
          
        A    & 2014-03-15 & 08:58:37-09:06:33   & CW    & --       & --  & --   & 0.0277                  & 6      & 0.044    & 23.4   & ~     \\
		  ~    & ~          & 09:22:02-11:10:14   & DD    & 0.1      & 2   & 7.5  & 0.037                   & 74     & 0.044    & 23.7     & ~  \\
          
        A    & 2014-03-16 & 08:27:48-08:41:16   & CW    & --       & --  & --   & 0.025                   & 9      & 0.048    & 27.8     & ~   \\
		  ~    & ~          & 08:49:44-10:59:44   & DD    & 0.1      & 2   & 7.5  & 0.075                   & 80     & 0.048    & 28.0     & ~  \\
          
        A    & 2014-03-17 & 08:12:30-08:18:33   & CW    & --       & --  & --   & 0.0222                  & 9      & 0.052    & 31.2    & ~   \\
		  ~    & ~          & 08:21:31-08:47:00   & DD    & 0.1      & 2   & 7.5  & 0.037                   & 15     & 0.053    & 31.3   & ~    \\ \hline
    \end{tabular}
\end{table*}

\begin{figure}
	\includegraphics[width=\columnwidth]{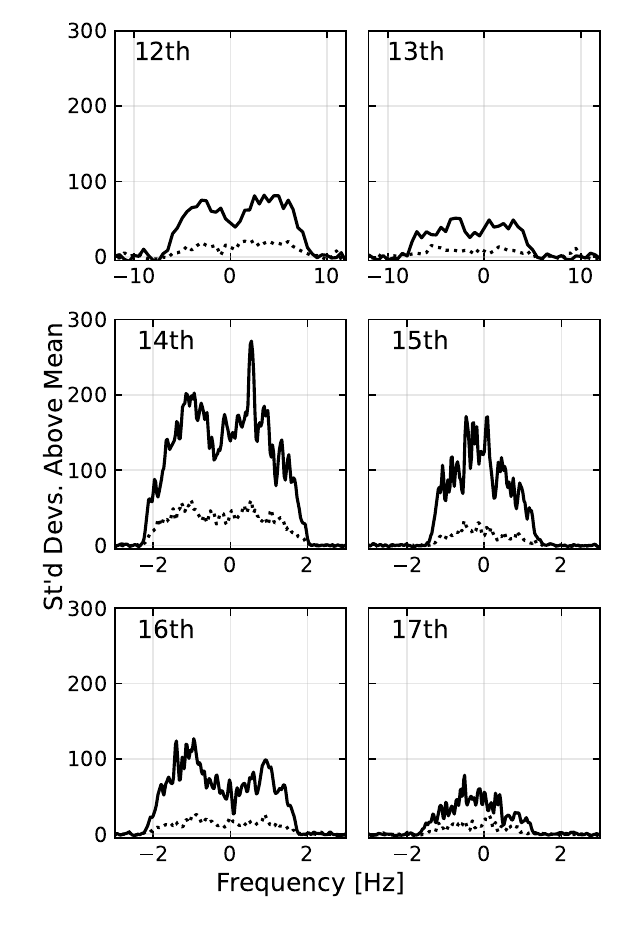}
    \caption{
    The summed CW spectra collected on each night between the 12th and 17th of March, 2014.
    The solid line shows the opposite-sense circularly polarised signal received, whilst the dotted line is the same-sense circularly polarised signal. 
    Observations on the 12th and 13th of March were taken at Goldstone, and those from the 14th-17th of March were collected with Arecibo.
    The resolution of each CW is described in Table~\ref{tab:RadarInfo}.
    }
    \label{fig:rs11_cw_plots}
\end{figure}

\begin{figure}
	\includegraphics[width=\columnwidth]{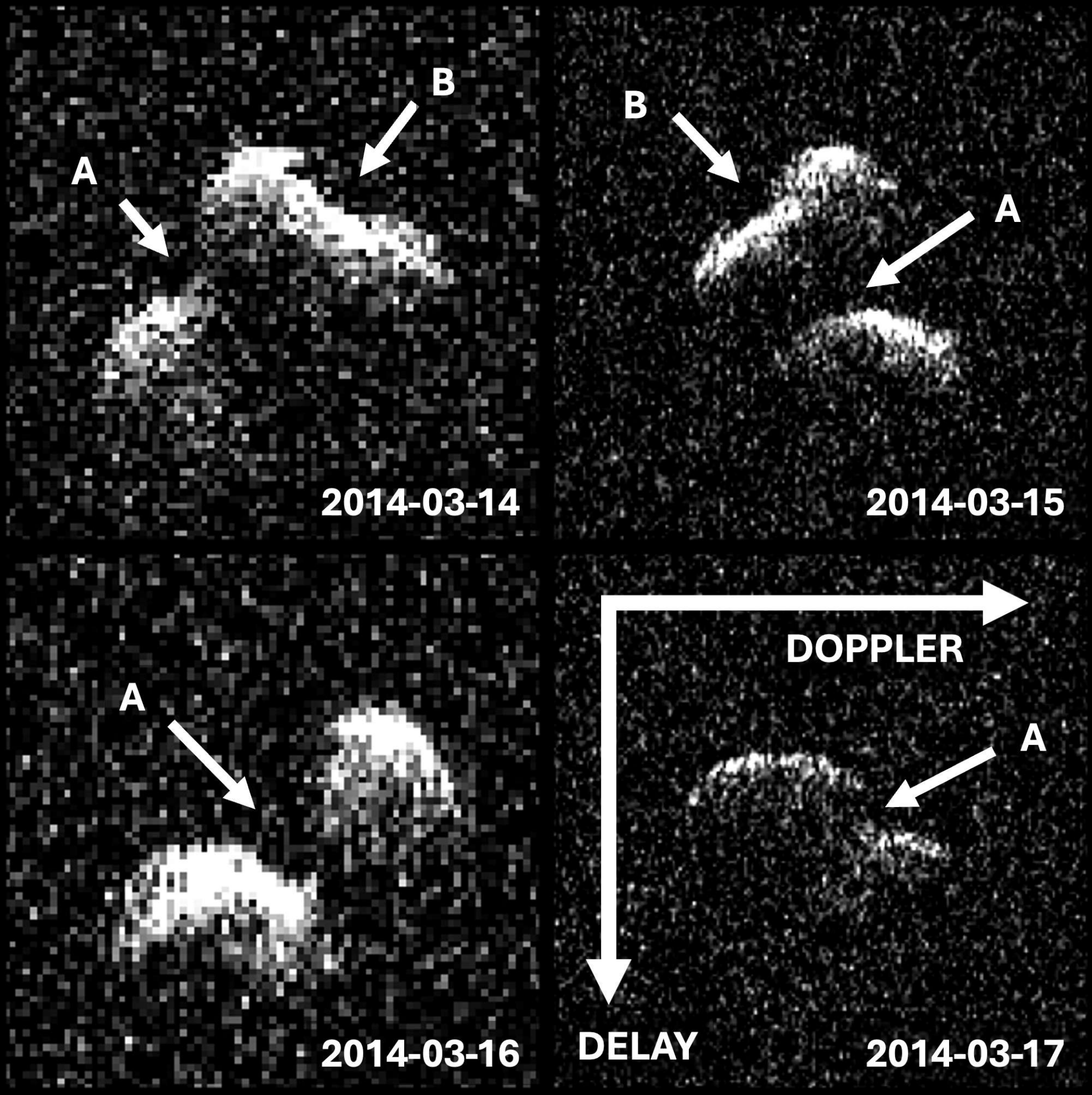}
    \caption{
    A collage of the first delay-Doppler for each night between 2014-03-14 and 2014-03-17 from Arecibo. 
    These images show the Doppler shift on the horizontal axis, increasing from left to right, and the time delay on the vertical axis, increasing from top to bottom.
    Markings on the images mark the location of (A) the concave neck structure between lobes, and (B) evidence for a crater or concavity on the larger lobe.
    The resolution for each day of observations is described in Table~\ref{tab:RadarInfo}.
    Here, each image is scaled and cropped so as to be $3$ Hz by $2~\mu s$ in size.}
    \label{fig:rs11_example_dd}
\end{figure}

To create our model, we used both optical lightcurves and radar observations. 
Both the radar data and the majority of the lightcurves were collected during the close approach in 2014, though we include an additional lightcurve collected in 2023, when \RS~was $0.78$ au from Earth.

Whilst new optical observations could be collected in 2027 to validate the spin state of this model, the next approach during which \RS~will be close enough to observe with current radar facilities and refine its shape will be in 2056, when it will approach within 0.020 au. 
This should provide detailed radar images with high signal-to-noise ratios.

\subsection{Optical lightcurves} \label{subsec:OpticalObservations}

Optical lightcurves are an important ingredient in creating a robust shape model and spin state. Whilst they contain limited information on the details of shape or concave features \citep{Harris&Warner2020_LC_CantTellBrickFromCB}, a contact binary shape still produces a lightcurve that differs from an elongated ellipsoidal shape, or its respective convex hull \citep{ShowalterEtAl2021_CBs-vs-Ellipsoids-TNO}, especially at higher phase angles \citep{Lacerda2007_CB-LCs-distinctive-high-phase-angle}.

Lightcurves provide a strong constraint on the object's rotational spin state. 
When combined with radar data, a strong constraint on the rotational period and pole allows the asteroid's size and shape to be modelled with greater confidence.

Optical data of \RS~used in this paper were predominantly collected in 2014 and published on ALCDEF \citep{WarnerEtAl2011_ALCDEFconcept} by \citet{Warner2014_Lightcurves_dp14_rs11} and \citet{Benishek2014_Lightcurves_rs11}. 
Additional lightcurves collected in 2023 would have provided additional constraints on the rotational pole and period due to the different viewing geometries; however, due to bad weather, only a quarter of a single rotation could be obtained, reducing its effectiveness.
Information for each of these lightcurves can be found in Table~\ref{tab:AllLightcurves}.

\subsubsection{Danish 1.54-metre telescope, La Silla -- 2023}

Located at an elevation of 2366 m at the La Silla Observatory in the Atacama desert, Chile, the 1.54-m Danish Telescope is operated jointly by the Niels Bohr Institute, University of Copenhagen, Denmark, and the Astronomical Institute of the Academy of Sciences of the Czech Republic. 
We collected one light curve of \RS~on the night of the $12^{\rm th}$ of July, 2023, in the Cousins R filter \citep{Bessel1990_UBVRI} with the Danish Faint Object Spectrograph and Camera (DFOSC) instrument which houses an e2v CCD 231 sensor with $2048 \times 2048$ square pixels and a $13.5 \times 13.5$ arcmin$^2$ field of view.

We reduced all images using standard flat-field and bias-frame correction techniques before median-stacking groups of 10 images to compensate for poor observing conditions. 
Aperture photometry was then performed on the sources using the average FWHM of every source in the frame.
Calibration of the lightcurve was performed following the procedure described in \citet{Kokotanekova2018_Thesis-OpticalReductionTechniques}.
Reference stars were cross-matched to the ATLAS-RefCat2 catalogue \citep{TonryEtAl2018_ATLASREFCAT2} using \texttt{calviacat} \citep{Kelley&Lister2019_CalviacatRef}.
The lightcurve was converted back to Cousins-R using the relation from \citet{Kostov&Bonev2018_Conversion-from-PanStarrs-to-Cousins}, and assuming a standard value of $(g-r) = 0.4\pm0.04$ for \RS.

Unfortunately, due to the short length of observing (approximately 25\% of \RS's full rotation) and poor conditions, the resulting lightcurve consists of only 6 stacked data points. 
These points, however, are enough for the brightness of \RS~to show a clear decrease of nearly 0.5 mag. 
Additional benefits of including this lightcurve in the modelling process come from its different phase angle and viewing geometry, as well as from its collection 9 years after the other lightcurves.

\subsubsection{Published Data}

Seven lightcurves of \RS~collected in 2014 greatly assisted the modelling process. 
These lightcurves, collected on $0.35$ m telescopes by both \citet{Warner2014_Lightcurves_dp14_rs11} and \citet{Benishek2014_Lightcurves_rs11} at the Palmer Divide Station in the US and the Sopot Astronomical Observatory in Serbia, respectively, provided strong coverage of multiple complete rotations of \RS.
These data are publicly available on the Asteroid Lightcurve Data Exchange Format (ALCDEF) database \citep{WarnerEtAl2011_ALCDEFconcept}.

\subsection{Planetary Radar} \label{subsec:Radar obs}

Described in more detail in \citet{VirkkiEtAl2023_RadarReview} and \citet{MagriEtAl2007_Betualia-Shape-Radar-Explanation}, radar observations are generated by emitting a signal of known properties and recording the reflected signal.
The strength of the measured radio wave is proportional to the inverse of the fourth power of distance, so it is difficult to observe any asteroids other than those very close to Earth, or the largest objects in the main belt.
For sub-km-sized near-Earth asteroids, this limiting distance is typically $< 0.05$ au for current facilities to obtain quality radar imaging, though, in the past Arecibo could observe objects farther from Earth. 
There are two types of radar observation: continuous wave spectra (CW), which record the power of the returning signal after it has been spread out over a range of Doppler shifted frequencies, and delay-Doppler images, which also embed into the radio wave a time-dependent code that allows measurement of the length of time the signal took to return to the observer.

The size of the object - specifically the diameter projected onto the plane of sky, $D$ - and the rotational period, $P$, is then estimated with the following equation:
\begin{equation}
    B = \frac{4 \pi D ~{\rm cos}\delta}{P\lambda},
\end{equation}
where $B$ is the bandwidth of the reflected signal, and $\delta$ is the sub-radar latitude.
Based on the CW data obtained, initial estimates of \RS's projected diameter ranged from $0.5-0.7$ km.
The CW data obtained can be seen in Figure~\ref{fig:rs11_cw_plots}.

Crucial for detailed shape modelling are delay-Doppler images.
In this paper, the time delay is plotted on the y-axis and the Doppler shift (relative to the ephemeris prediction) on the x-axis.
Examples of delay-Doppler images collected are shown in Figure~\ref{fig:rs11_example_dd}.
These provide fine detail of the surface structure and shape down to $7.5$ m.

We obtained a mixture of both CW and delay-Doppler observations from Goldstone and Arecibo between the $\rm12^{th}$ and $\rm17^{th}$ of March, 2014.
Described fully in Table~\ref{tab:RadarInfo}, observations consist of transmitting $8560$ MHz ($3.5$ cm wavelength) radio waves from Goldstone, or $2380$ MHz ($12.6$ cm wavelength) radio waves from Arecibo, and recording the reflected signal.
The best radar observations were collected with Arecibo while \RS~was already travelling away from Earth.
These delay-Doppler images have delay resolutions of $7.5$ m and Doppler resolutions of either $0.074$ or $0.037$ Hz, whilst the earlier Goldstone observations have a resolution of $\rm 0.48~Hz \times 37.5~m$.

When inspecting both the delay-Doppler images and continuous wave spectra, a clear bifurcation in the signal is observed, requiring an irregular bilobate shape not typical of other contact binary asteroids.
Additionally, we observed evidence of a concavity on one of the lobes, highlighted in the upper two images in Figure~\ref{fig:rs11_example_dd}.

All of the CWs and delay-Doppler images were then summed into groups to increase the SNR.
Though sums were kept to a maximum of $6\%~(20^{\circ})$ of a full rotation, the elongation of the object results in a maximum rotational smearing of the furthest extents of the target of up to $\sim 125$ m, or $\sim16$ pixels, when viewing the long axis of \RS.
In practice, a combination of slightly oblique viewing geometries and smaller sums limit the blurring to less than the theoretical maximum, and this effect can be accounted for in the modelling process.

\section{Modelling} \label{sec:Modelling}

As the existing lightcurves collected in 2014 already provided a reasonable constraint on the period, we started the modelling process directly with the radar data using the \shape~modelling software \citep{MagriEtAl2007_Betualia-Shape-Radar-Explanation}. 
From here, we follow the same principles of the modelling process described in \citet{CannonEtAl2025_DP14shape}, starting with a simplified model to perform a pole scan, then increasing complexity with vertex models.
To account for the rotational effects introduced by summing the images, we used \shape's smearing parameter to render artificial delay-Doppler frames as unweighted means of individual `views', each view corresponding to a raw delay-Doppler image in the sum.

\subsection{Pole scan with spherical harmonics} \label{subsec:PS with spherical harmonics}

\begin{figure}
	\includegraphics[width=\columnwidth]{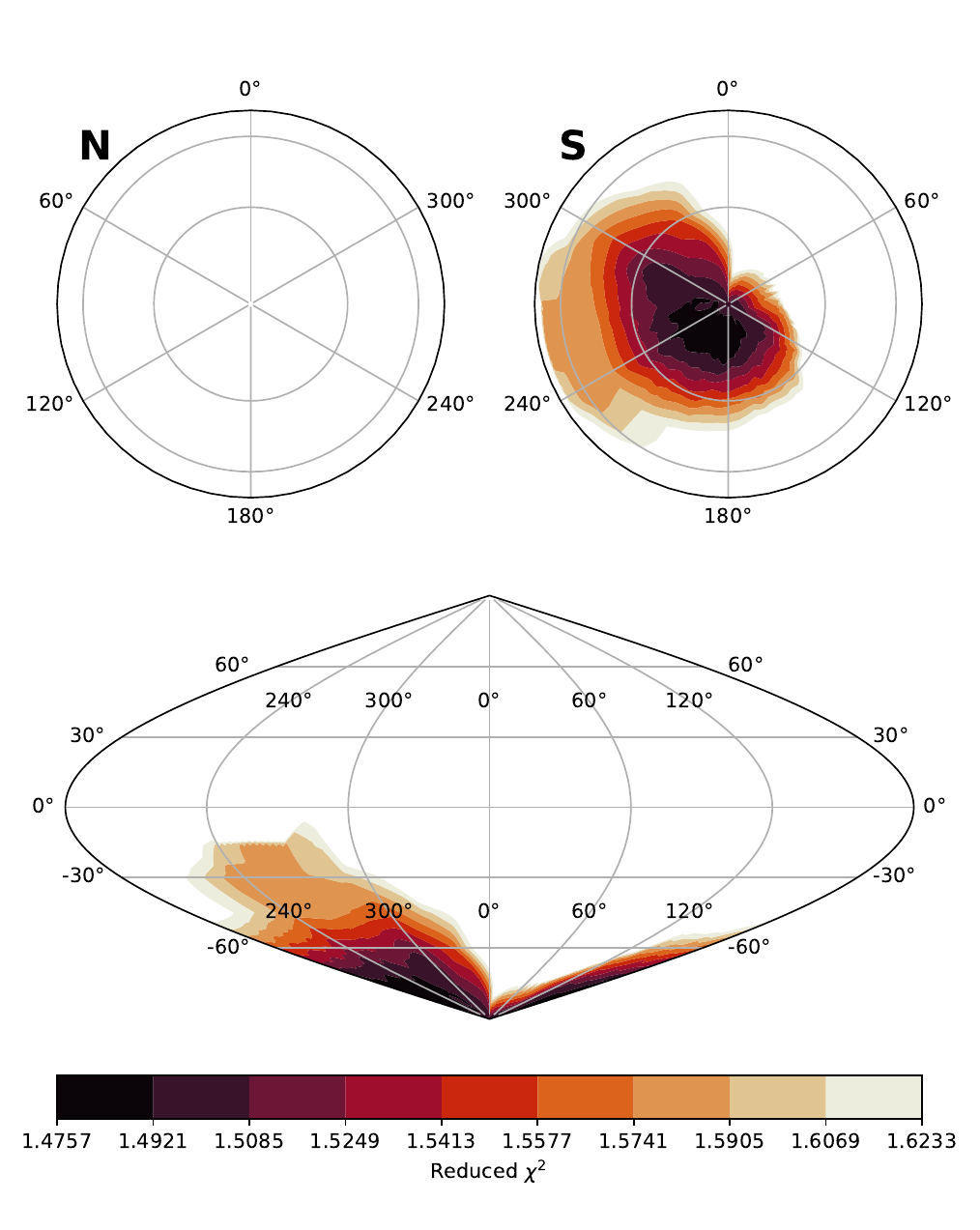}
    \caption{Pole scan generated using a spherical harmonic model to approximate the shape of \RS~using \shape. The scan used all available radar and lightcurve data, with the initial conditions set to a spherical-harmonic description of a sphere with a radius estimated from the radar images. For each data point, all shape parameters and the period were allowed to vary, but the rotational pole orientation was kept constant. Results are shown in ecliptic coordinates, with darker areas indicating better fits. White areas have a reduced $\chi^2$ above 10\% of the best value.}
    \label{fig:radar_polescan}
\end{figure}

To model the rotational pole of \RS, we initially employed a bi-ellipsoid approach to approximate the contact binary shape \citep[e.g.][and others]{MagriEtAl2011_1996HW1-shape-CB, RozekEtAl2019_JV6-shape, BrozovicEtAl2018_ApophisRadarObsAndModel}.
However, due to the concavity in the larger lobe, we struggled to obtain consistent results and proceeded with a spherical harmonic model, which produced more stable fits across neighbouring pole solutions.
The spherical harmonic model used was a single-component 4th-order model.
Tests at different orders of spherical harmonic expansion revealed that the 4th order was able to approximate the shape best, whilst keeping computing requirements low enough to be feasible.

Spherical harmonics, while used in the early cases of modelling with \shape~when all free parameters were modelled at once \citep{Hudson&Ostro1994_Castalia-Shape}, are no longer used as frequently to model contact binaries due to the increasing complexity of the models and software.
The modern implementation of \shape~now fits the position of each vertex or spherical harmonic coefficient as a separate parameters \citep{MagriEtAl2007_Betualia-Shape-Radar-Explanation}.
One drawback of spherical harmonics is the requirement for star-shaped geometry - a shape for which every line drawn from the centre of mass to any point on the surface must lie entirely within the body of the shape.
Whilst this is the case for \RS, contact binaries with strong neck features cannot be approximated with a single spherical-harmonic component.

The most significant advantage of spherical harmonic models is that it is possible to simulate the observed data for irregular shapes with a higher degree of accuracy than ellipsoids, while maintaining few free parameters to describe the shape.
On the other hand, the `global' nature of spherical harmonics makes it exponentially more challenging to create small-scale features when varying only one parameter at a time.
Therefore, spherical harmonics are ideal for the low-resolution pole scan stage in the modelling process.

Another challenge with spherical harmonics can be creating a good initial model.
Due to their global nature, it is much less intuitive to create reliable spherical-harmonic models by hand than manipulating the size and position of ellipsoids.
Therefore, we started each pole scan with a sphere, with all other spherical-harmonic components set to 0.
However, due to the elongated nature of the object in the data, \shape~is quite reliant on the initial rotation phase of the sphere.
If the initial rotation is suboptimal, \shape~can, as early as the dipole coefficients, deviate from the best solution by stretching the model in the wrong direction, subsequently getting trapped in a local minima with a poor solution.
To counter this, we ran parallel pole scans with different initial rotation phases of the spheres, allowing the object's rotation to vary while the pole position remained fixed.
The 3-dimensional output is then flattened into a 2-dimensional pole scan by selecting the best solution found for each fixed pole direction.

With this method, we recorded the quality of fit for a range of pole solutions across the celestial sphere, increasing the density of tested points in regions where the quality of the solution increased.
Displayed in Figure~\ref{fig:radar_polescan}, the best solution at this stage was found to be a rotational pole of $(\lambda=200^{\circ}, \beta=-85^{\circ})$.
Already at this stage, the shape visually matched the delay-Doppler imaging well, although it did not sufficiently replicate any concavities such as on the larger lobe, or in the neck area.

\subsection{Vertex modelling}

Following the pole search, we selected the best solutions to continue refining as vertex models.
Initially, low-resolution models with 500 vertices were used and run many times with different input parameters and penalty functions.
Before each fit, the vertices' order was randomised to ensure that no artefacts were introduced by fitting them consecutively.

Penalty functions are required to modify the $\chi^2$ of the output so that the user can discourage non-physical solutions.
Example penalty functions include those that discourage non-principal-axis rotation or force the centre of mass to lie on the rotational axis.
Additionally, penalty functions exist for `smoothing' the surface, so that a single vertex is not brought far from the body to fit noise outside the signal of the target.
For \RS, we used this smoothing penalty and did not penalise concavities specifically, given the strong concave features in the data.
Tests with penalties against concave features showed an apparent discrepancy with the data, preventing the concave neck feature from replicating the dramatic, steep slopes observed in the data.

\begin{figure*}
	\includegraphics[width=\textwidth]{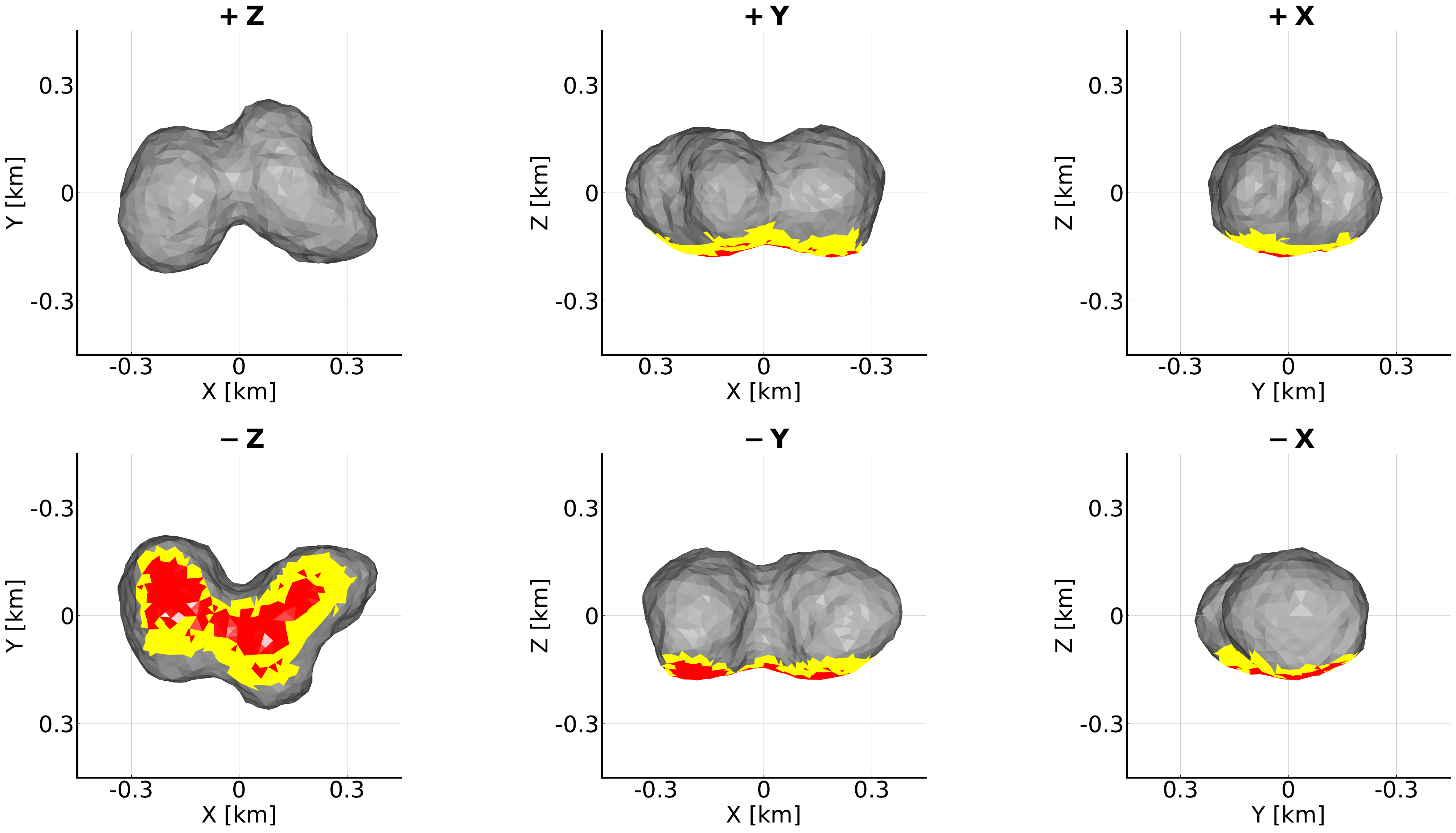}
    \caption{The six principal axis views of the final shape model of \RS~generated with both optical and radar data in \shape. The model has pole solution $\lambda = 225^{\circ}$ and $\beta = -80^{\circ}$. The average length between the $1500$ vertices is $26.9$ metres. Red shading is applied to facets not viewed by the delay-Doppler radar imaging, while yellow shading is applied to facets viewed only at scattering angles greater than $60^\circ$ in the delay-Doppler images.}
    \label{fig:Radar_Model}
\end{figure*}

After 500 vertices, the best 10 models were advanced to 1500 vertices and modelled further with different vertex orders. 
Subsequently, the 5 best models were rerun again with shuffled vertices to test for further improvements and ensure no artifacts had been introduced.
For these models, the delay-Doppler contribution to $\chi^2$ was above $80\%$, as the spin-state had already been determined and the rotational pole was fixed at this stage.
To ensure the period was consistent with the lightcurve data, its $\chi^2$ contribution was kept to $\sim 10\%$.
The model with the best $\chi^2$ is shown in Figure~\ref{fig:Radar_Model} and had a pole solution $(\lambda = 225^{\circ},~\beta = -80^{\circ})$ and a period of $P = 4.445$ hours.

With $1500$ vertices, the minimum length between two vertices was $8.5$ m, and the mean length was $26.9$ m.
When comparing this with the $7.5$ m resolution of the best-quality radar data, we ensure that no over-fitting to noise in the data or artefacts caused by summing images.

A comparison of this model to the observations is in Appendix~\ref{app:model fits}.
The model echoes reproduce the delay-Doppler images well, replicating both the relative lobe sizes, but also the neck and large concavity.
The CW data has a poorer fit than the imaging but matches the bandwidth of the simulated echo.
Discrepancies around the peaks of the CW spectra may be due to either noise in the data, or the cosine scattering law used to simulate the radar observations being insufficient.
The lightcurves, meanwhile, show good fit to the data in both depth of the large minima and smaller variations across the rotation period.

As the only lightcurve in 2023 spans only 25\% of the rotation period, it does not provide a strong constraint on the period as might be hoped from a second epoch of observations.
Many aliased period solutions could fit all the lightcurves, and this is reflected within the range of periods output from the models in the pole scan.
We tested the final model with a grid of different rotational periods and scaling factors to ensure that the solution reached was the best solution.
However, no newly tested model was better than the initial starting point, implying that \shape~had already reached an optimal period solution for this pole orientation and morphology.

Radar images provided good coverage of the surface, with red shading applied to facets not viewed by the delay-Doppler radar imaging in Figure~\ref{fig:Radar_Model}. 
Yellow shading has been applied to facets viewed only at scattering angles greater than $60^\circ$.
These are areas where we lack detailed shape information from the radar delay-Doppler images. 
Still, the shape is fit against the lightcurves, and to best balance the centre of mass and rotational axes of the model, to ensure a physical solution.

Although few penalties against concavities were used to create the final shape model, prior tests with strong concavity and smoothing penalties also reproduced the strong concavities, albeit with noticeably shallower gradients that did not match the steep slopes seen in the data.
Therefore, these concavities are unlikely to be artefacts of data summation or caused by over-fitting to noise.

With the shape finalised, we looked to constrain the radar scattering exponents $n_S$ and $n_X$, for S- and X- band radar observations respectively.
We ran a grid search between $1 < n < 6$ on the scattering exponents for both the Arecibo and Goldstone delay-Doppler images.
We find a value of $n_S = 1.6^{+0.7}_{-0.5}$, where errors are conservatively derived from fits within $0.1\%$ of the best solution.
Due to the low SNR and fewer images from Goldstone, we were not able to constrain a value for scattering law exponent $n_X$.

\section{Discussion} \label{sec:Discussion}

\subsection{The shape and spin state of 2000 \RS} \label{subsec:rs11shape}

\begin{table}
    \centering
    \caption{Properties of (275677) 2000 \RS~derived from the final shape model created using both radar and optical observations. 
    Uncertainties on shape and spin state are conservative estimates derived from statistical analysis of the spherical harmonic pole scan models with a $\chi^2$ within $5\%$ of the best solution. 
    Uncertainties on radar scattering exponent $n_S$ are derived from a grid scan ranging from $1 < n < 6$.}
    \label{tab:rs11_properties}
    \begin{tabular}{lcr} \hline \hline 
    Parameter               & Value     & Uncertainty  \\ \hline
    $D_{eq}$ [km]           & 0.471     & 0.085 \\
    Volume [km$^3$]         & 0.055     & 0.035 \\
    Surface Area [km$^2$]   & 0.846     & 0.309 \\ \hline
    Physical Extents [km]   &           &  \\
    ~~~~X                   & 0.72      & 0.09 \\
    ~~~~Y                   & 0.48      & 0.14 \\
    ~~~~Z                   & 0.37      & 0.05 \\
    DEEVE Extents [km] \& Ratios &      &  \\
    ~~~~2a                  & 0.74      & 0.10 \\
    ~~~~2b                  & 0.41      & 0.07 \\
    ~~~~2c                  & 0.34      & 0.05 \\ 
    ~~~~a/b                 & 1.80      & 0.07 \\
    ~~~~b/c                 & 1.21      & 0.15 \\ \hline
    Spin state              &      & \\
    ~~~~P [h]               & 4.445     & 0.001 \\
    ~~~~$\lambda~[^\circ]$  & 225       & 80 \\
    ~~~~$\beta~[^\circ]$    & -80       & 9  \\ \hline
    Radar Scattering Exponent & & \\
    ~~~~$n_S$               & 1.6       & $^{+0.7}_{-0.5}$ \\ \hline    \end{tabular}
\end{table} 

The final shape model (Figure~\ref{fig:Radar_Model}) has a pole of $\lambda = (225\pm80)^{\circ}$ and $\beta = (-80\pm9)^{\circ}$ and a rotation period of $P = (4.445\pm0.001)$ h, in agreement with the previous literature value of $(4.444\pm0.001)$ h.
This rotational period places \RS~as the second fastest rotator in the modelled contact binary population, behind only Castalia.
When comparing to a wider sample of contact binary candidates that have been observed with radar but not yet modelled, \RS~has a period within the $15\%$ fastest (\url{https://echo.jpl.nasa.gov/History/}, accessed 20/02/2026). 
A rotational pole in the southern hemisphere places \RS~amongst the majority of NEAs.
This preference for retrograde rotators in NEAs is due to the transport of retrograde rotators from the main asteroid belt into certain resonances by the Yarkovsky effect, which in turn feed the NEA population \citep{Vokrouhlicky2998_YarkovskyDriftInMBA, BottkeEtAl2002_NEOsSuppliedByMBA, LaSpinaEtAl2004_RetrogradeNEAsFromYarkovsky}, and the YORP effect preferentially pushing rotational poles to the north and south poles \citep{Rubincam2000_YORP-concept, VokrouhlickyEtAl2003_YORP_Editing_Spin_States}.
During the polescan stage of the modelling process, we found a range of periods spanning from 4.443 h to 4.447 h in our final models due to the little information provided by the short 2023 lightcurve.
Therefore, our constraint on the rotational period is not any stronger than prior estimates from lightcurves alone.
This will not dramatically affect the resulting shape of \RS~-- all pole solutions from the pole scan, irrespective of period, had a shape similar to our final solution.

Other properties of its physical shape and spin state are described in Table~\ref{tab:rs11_properties}. 
Uncertainties given for all properties were conservatively estimated by fitting skewed normal distributions to the parameters of pole scan models with a reduced $\chi^2$ within $5\%$ of the best solution.
$5\%$ was chosen as a cut-off for which \shape~found similar shapes that visually fit both the lightcurves and delay-Doppler images moderately well.
Outside this limit, the spherical-harmonic nature of the models led to more irregular shapes with noticeably poorer fits to the data.
The uncertainty in the Z-axis of this model is likely underestimated as the southern polar region of the object was not imaged with radar.

\RS's shape is unusual compared to previously modelled contact binaries, with the two lobes being of nearly equal volume ($m_1/m_2 \approx 0.86$), compared to most other modelled contact binaries, which have volume ratios $\sim0.4-0.6$ (calculated from shape models listed in \citet{CannonEtAl2025_DP14shape}).
Its larger lobe is an irregular `bean'-shaped lobe with a large concavity on the outward-facing side, opposite the neck, whilst its smaller lobe is more spherical with no significant surface features.
When calculating the alignment of the two lobes, using the method laid out in \citet{WimarssonEtAl_AngleOfCBs}, we calculate that the smaller lobe is placed at an angle of $\sim 59^{\circ}$ degrees to the longest axis of the main lobe.
This is unusual compared to most other radar contact binaries, which have angles below $25^{\circ}$ (only (4486) Mithra - an NEO - and comet 67P also have values comparable to \RS).
These unusual locations of the small lobe may point to an unusual evolutionary past, though from the ground-based shape model alone, this remains speculation.
It could be inferred from the large concavity and the high volume ratio between the two lobes that the larger lobe initially was significantly larger and better aligned with the smaller lobe, bringing it in line with other modelled contact binaries.
However, for a volume ratio of $0.5$, the larger lobe would be required to be $\sim 70\%$ more massive than currently observed (assuming a uniform density throughout) prior to any impact/disruption event.
Such a large disruption event that doesn't seem to have affected the shape of the smaller lobe seems unlikely, so this could also imply the concavity formed before the contact binary nature of \RS.

With the discovery of Selam, there are renewed theories on whether contact binaries could form in debris disks around larger primary objects.
Resulting simulations show that these chaotic systems could form contact binaries \citep{WimarssonEtAl_AngleOfCBs}, and the shape of Selam has been recreated in simulations by the subsequent collisions of 4 separate moonlets \citep{RaducanEtAl2025_SelamFormationFromMoonlets}.
It would therefore not be unreasonable to expect \RS~to have formed in a chaotic environment through several consecutive collisions, creating the large concavity on the larger lobe.
In fact, it is common for contact binary objects to have craters and concavities on the surface that do not appear related to the neck structure, with \DP, \JV, \HW, and Toutatis all displaying evidence of impact craters \citep{CannonEtAl2025_DP14shape, RozekEtAl2019_JV6-shape, MagriEtAl2011_1996HW1-shape-CB, ZhuEtAl2014_Toutatis-ChangE2-Shape}.
Other contact binaries, such as (68446) 2006 \KZ, Mithra, and (1981) Midas, also display concave features \citep{ZegmottEtAl2021_KZ66-YORP-Detection-and-Shape, BrozovicEtAl2010_Mithra-Shape-Model, McGlassonEtAl2022_MidasShapeModel}.

Interestingly, the two objects with similar lobe orientations to \RS, Mithra and 67P, both feature concavities on the outward-facing sides of their lobes.
Whilst 67P's morphology is likely to be shaped in part due to surface activity \citep{ElMaarryEtAl2019_Review_67P-morphology-surface-evolution, Rezac&Zhao2025_Activity-causing-small-scale-concavities-67P}, it is also possible that its bilobate nucleus was formed from a non-catastrophic collision of two bodies \citep{Jutzi&Asphaug2015_Low-Velocity-CB-formation-comets, Jutzi&Benz2017_Low-Velocity-CB-formation-comets_2} in a similar way to how we might expect the contact binary asteroids to form.
These simulations, however, have not yet been able to replicate these outer concavities.

Alternate theories for \RS's formation could be that it formed from a combination of either YORP or BYORP \citep{Rubincam2000_YORP-concept, Cuk&Burns2005_BYORP-concept}.
By speeding up the rotation of a parent body, the YORP effect could cause loose regolith to flow across the surface and create a contact-binary-like shape \citep{Sanchez&Scheeres2018_SpinUp_Creating_Itokawa-like_shape} or, if the body is rigid, fracture into two lobes \citep{Scheeres2007_RotationalFissionOfCBs}.
Currently, there are 2 known contact binaries with a YORP detection: Itokawa and \KZ~\citep{LowryEtAl2014_Itokawa-YORP-and-densities, ZegmottEtAl2021_KZ66-YORP-Detection-and-Shape}.
Once in a stable binary system, formed from a rotational fissure or debris disk, the BYORP effect could cause a slow in-spiral of the lobes - gently merging to create a contact binary shape \citep{Jacobson&Scheeres2011_CB-formation-from-BYORP}.
Whilst this is exact scenario is unlikely to apply to \RS~due to its larger lobe angle, most modelled contact binaries have lobe angles between $10$ and $20$ degrees, which could indicate formation from a binary system, such that the smaller lobe makes contact with the long axis of the elongated lobe, before experiencing a drag force and coming to rest slightly offset from the longer axis.

\subsection{Disk-integrated properties}

\begin{table}
    \centering
    \caption{Disk-integrated properties of (275677) 2000 \RS~from the collected radar CW spectra. Displayed are the OC radar cross-section, $\sigma_{\rm OC}$, the circular polarisation ratio, CPR, calculated from $\sigma_{\rm OC}/\sigma_{\rm SC}$, the projected areas of \RS~at the time of observation, estimated using \shape, and the albedo, $\eta_{\rm OC}$, which is the ratio between the radar cross section and the projected area.}
    \label{tab:cw_integrated_properties}
    \begin{tabular}{cccccc} \hline \hline 
    Tel. & Date       & $\sigma_{\rm OC}$ & CPR             & $\rm A_{proj}$ & $\eta_{\rm OC}$ \\
    ~    & ~          & [$\rm km^{2}$]    &                 & [$\rm km^{2}$] &                 \\ \hline
	  G    & 2014-03-12 & $0.032 \pm 0.008$ & $0.22 \pm 0.08$ & $0.204$        & $0.16$          \\
    G    & 2014-03-13 & $0.027 \pm 0.007$ & $0.23 \pm 0.08$ & $0.193$        & $0.14$          \\
    A    & 2014-03-14 & $0.015 \pm 0.004$ & $0.26 \pm 0.09$ & $0.212$        & $0.08$          \\
    A    & 2014-03-15 & $0.017 \pm 0.004$ & $0.16 \pm 0.06$ & $0.180$        & $0.09$          \\
    A    & 2014-03-16 & $0.019 \pm 0.005$ & $0.20 \pm 0.07$ & $0.215$        & $0.09$          \\
    A    & 2014-03-17 & $0.014 \pm 0.004$ & $0.27 \pm 0.09$ & $0.193$        & $0.08$          \\ \hline
    \end{tabular}
\end{table}

Using the collected CW data, we can obtain disk-integrated properties of \RS, which provide insights into its composition and surface properties.

As radar observations are performed by emitting circularly polarised radio waves, we measure the polarisation of the returning wave and record the ratio of the same- and opposite-circular (SC and OC) polarisation radar cross sections.
The circular polarisation ratio (CPR), defined as $\sigma_{SC}/\sigma_{OC}$, can be related to the spectral classification of an asteroid and is a measure of the size, texture, and shape of particles on the surface of the target \citep{OstroEtAl1985_DensityFromRadarAlbedo, VirkkiEtAl2014_SpectralClassificationFromSCOC+RadAlbedo, HicksonEtAl2021_UnknownShapeEffectsOnSCOC, RiveraValentinEtAl2024_SpectralClassificationFromSCOC}.
We find \RS's CPR to be between $0.16$ and $0.27$ (Table~\ref{tab:cw_integrated_properties}).
Variations in these ratios may be caused by either varying viewing geometries or different sections of the surface being visible during different observations.
The CPR can depend on both the material and size of regolith on the surface, so it is possible that different lobes have different structure or material properties \citep{VirkkiEtAl2013_SCOC_theory}.
Unfortunately, it is difficult to distinguish whether it is the size distribution or the material of the surface causing these variations.
The average CPRs for Goldstone and Arecibo respectively are $0.225$ and $0.223$.

We find that 4 of the CW observations, collected on the 12th, 13th, 14th and 16th, clearly observe both lobes; however, CW observations on the 17th ($0.27$) and the 15th ($0.16$) viewed predominantly only the larger and smaller lobes, respectively.
The plane-of-sky visualisations of the model for each CW echo power spectrum are displayed adjacent to the model fit in Appendix~\ref{figapp:radar_cw_fit}.
Such a CPR places \RS~with other S-complex NEAs, which have a mean ratio of $0.26\pm0.08$ \citep{RiveraValentinEtAl2024_SpectralClassificationFromSCOC}.

We estimate the opposite-sense radar albedo, $\eta_{\rm OC} = \sigma_{\rm OC}/A_{proj}$ using the radar cross-section of each CW and the projected area at the time of observation (Table ~\ref{tab:cw_integrated_properties}.
We find that Goldstone data ($\sim3.5$ cm) gives albedos around $0.15$, whilst Arecibo ($\sim 12.6$ cm) estimates albedos to be $\sim0.08$.
This discrepancy could be attributed to different scattering properties between the longer S-band and shorter X-band wavelengths.
It is not unusual to obtain differing radar albedos from these two observatories \citep[see][]{BrozovicEtAl2024_2018EB_shape, BrozovicEtAl2018_ApophisRadarObsAndModel, LawrenceEtAl2018_RadarObs2005wc1}, though often these values are within $1\sigma$ of each other.
When comparing to other asteroids, we find that \RS~has a lower albedo than average - \citet{VirkkiEtAl2022_30percent-CB} reports a mean opposite-sense albedo for NEOs of $0.21 \pm 0.11$, whilst a smaller selection of S-type asteroids have a mean of $0.19 \pm 0.06$.
However, such an albedo places \RS~in a similar position to two other radar modelled contact binaries S-complex asteroids: Itokawa ($0.138 \pm 0.014$) and \HW~($0.15 \pm 0.04$).

We can estimate the taxonomic class of \RS~using the radar albedo and SC/OC ratio \citep{VirkkiEtAl2014_SpectralClassificationFromSCOC+RadAlbedo}.
Although multiple taxonomic classes are consistent with the measured radar properties, \RS~is consistent with the S-complex group, in line with its Sa-type classification.

Finally, we use the final shape model to estimate a geometric albedo of \RS~with 
\begin{equation}
    p_v = \left({1329}/{D}\right)^{2} \times10^{-0.4H}
\end{equation} 
\citep{Harris1997_OpticalAlbedo}.
We use the equivalent diameter, $D_{eq}=(0.471\pm0.085)$ km and a value of $H = 19.22$ from Horizons to estimate an albedo of $p_v = 0.16 \pm 0.06$.
The previous estimate of $p_v = 0.35^{+0.24}_{-0.14}$ was derived from infrared observations assuming a spherical shape \citep{ThomasEtAl2011_SpectralClassificationVsAlbedo}.
Though there is less information on the albedos of Sa-type asteroids, this new value is in better agreement with geometric albedos of other S-type asteroids \citep[Fig. 9 in][]{Ryan&Woodward2010_Albedos_to_Spectral_Type}.

\subsection{Comparison to the previous model} \label{subsec: Comparison to previous model}

The shape model presented in this paper bears some resemblance to the previous shape model created in 2014 \citep{BrauerEtAl2014_ConfAbs2000rs11_shape}.
Displayed in Appendix~\ref{app:2014 model} along with its fits to the lightcurves, the older model has a much flatter, larger lobe than the `bean' shape in the current model.

The spin state of the new model is also significantly different.
Whilst the period remains very similar (as the prior period was taken from the 2014 optical data), by integrating the optical and radar data, the rotational pole is now constrained to a region closer to the southern pole.
The previous pole solutions were $(155^{\circ},30^{\circ})$ or $(335^{\circ},-30^{\circ})$, while the new pole solution is $(225^{\circ},-80^{\circ})$.
These differences arise from the use of only radar delay-Doppler images in previous modelling efforts, and are immediately evident when comparing the model to the lightcurves.
We found it impossible to fit the northern hemisphere solution such that the model was in phase with both the radar and lightcurve observations simultaneously (Appendix~\ref{figapp:2014np_lc_fit}).
For the southern hemisphere solutions, the simulated observations are in phase with both modes of data; however, the lightcurve amplitudes are not in agreement (Appendix~\ref{figapp:2014sp_lc_fit}), indicating that the shape and pole still needed further refinement.

Finally, we can inspect the model relative to the CW radar data. Whilst, in theory, this should contain similar information to the delay-Doppler images, the `flattened' nature of the data allows for higher SNR and a more obvious mismatch with the data. 
Here, we find that the 2014 model's asymmetric lobes do not match the equally distributed signal observed in the data.

This highlights the importance and need for including multiple data sources in modelling.
Whilst delay-Doppler images can contain incredibly high-resolution shape information ($7.5$ m in this case), the nature of the data allows for degenerate solutions and for discrepancies with the data to be more easily overlooked.
Including the lightcurves immediately points to a rotational pole near the southern pole, whilst the CW data better highlight the nature of the relative sizes of the two lobes than is evident in any delay-Doppler imaging.

\subsection{Gravitational Environment}

\begin{figure*}

\begin{subfigure}{\textwidth}
	\includegraphics[width=0.95\textwidth]{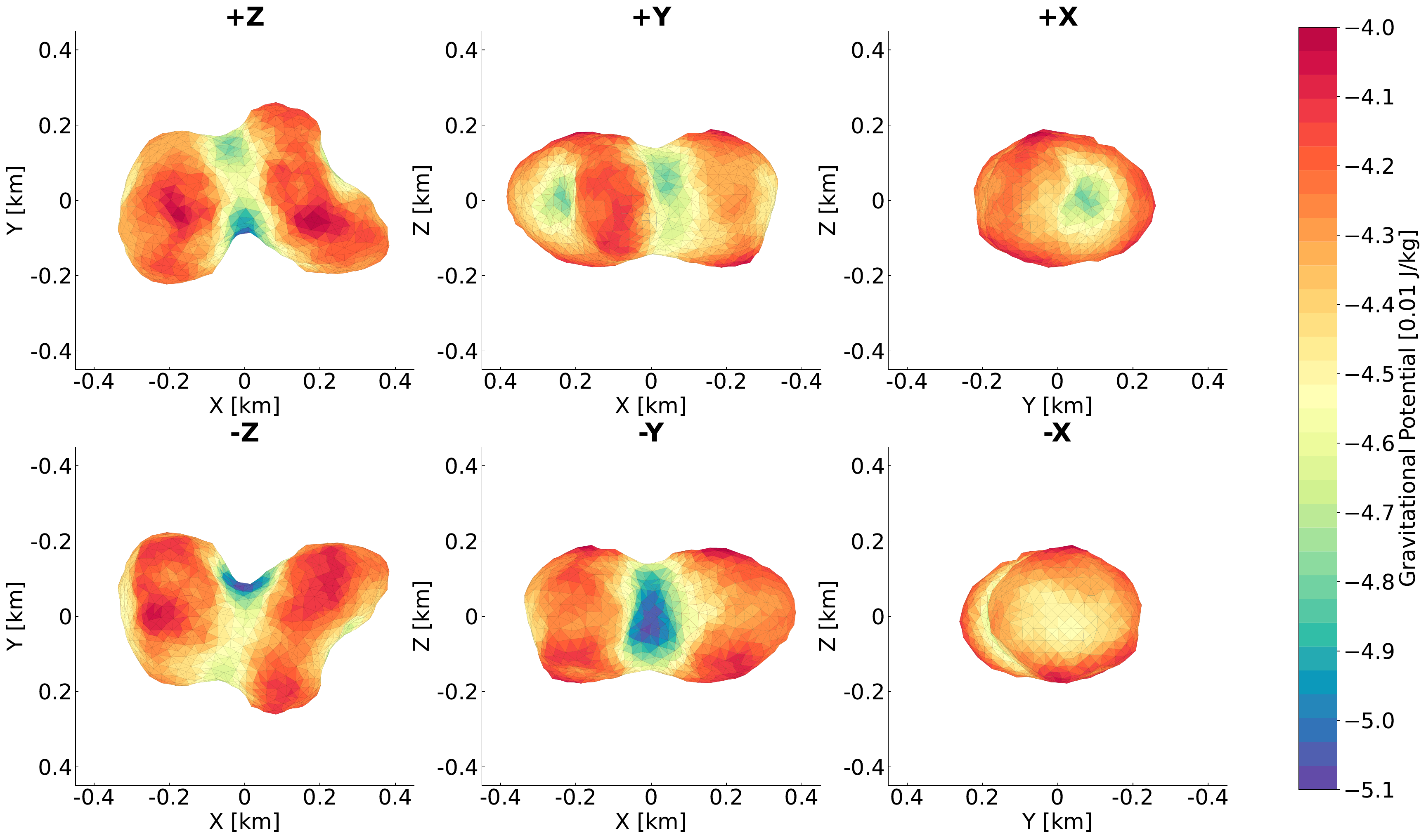}
    \caption{~}
    \label{fig:GravPotential}

\end{subfigure} \\ \vfill

\begin{subfigure}{\textwidth}
	\includegraphics[width=0.95\textwidth]{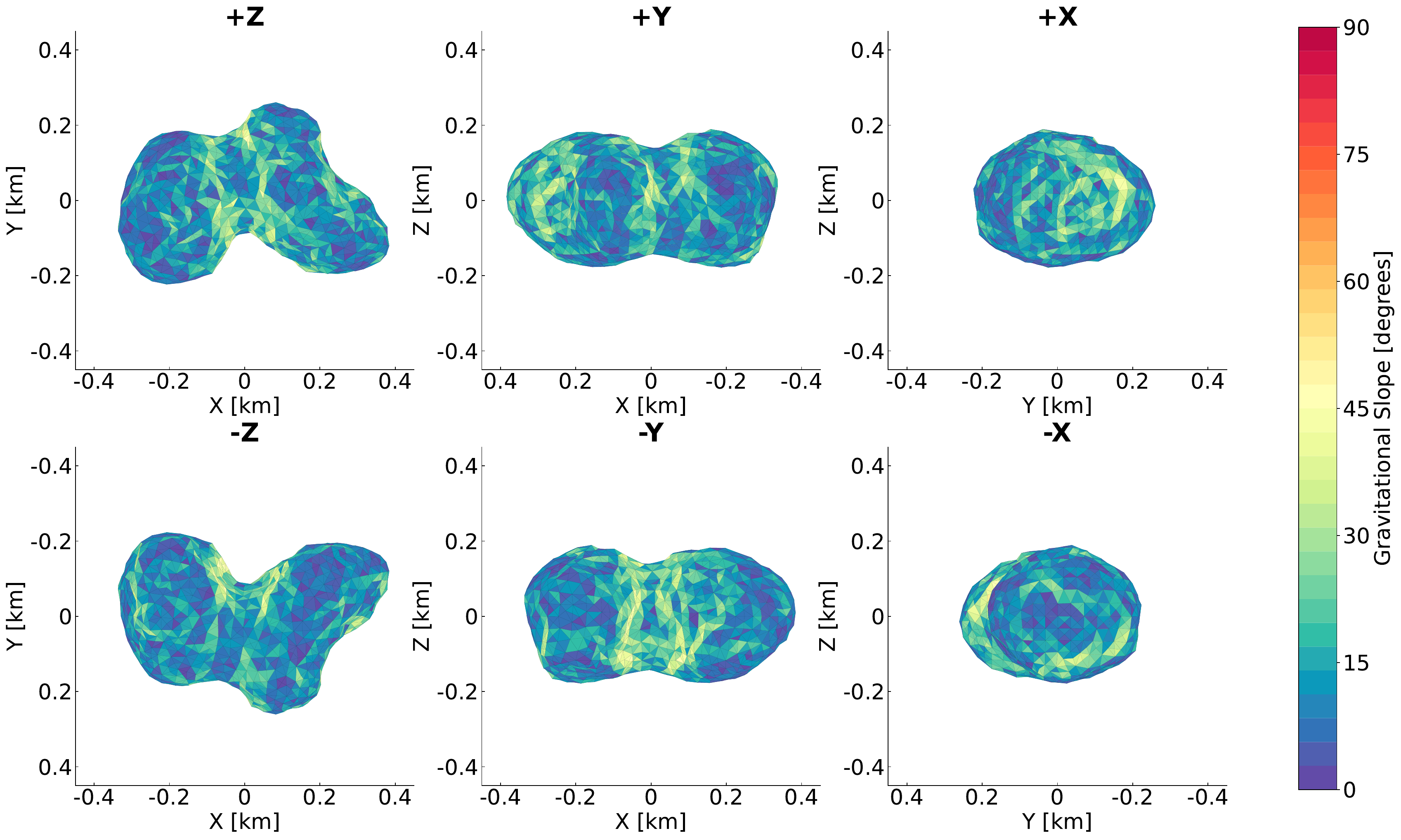}
    \caption{~}
    \label{fig:GravSlopes}

\end{subfigure}
\label{fig:GravInfoFigs}
\caption{The gravitational potential (\ref{fig:GravPotential}) and gravitational slopes (\ref{fig:GravSlopes}) of \RS, assuming a uniform density of $\rm 2.66~g.cm^{-3}$. 
Gravitational slopes are given relative to the normal of each facet, such that an angle of $0^{\circ}$ points into the surface, and any angle greater than $90^{\circ}$ would cause regolith to separate from the surface. 
The six plots show the model in both directions along the X, Y, and Z axes.}
\end{figure*}

As an Sa-type asteroid in the Bus-DeMeo taxonomical system \citep{LazzarinEtAl2004_RS11-spectral-type, DeMeoEtAl2009_BusDeMeo-system}, \RS~is between the S- and A-type classifications, redder than normal S-types.
However, due to the smaller number of Sa-types or A-types, population information for densities or colours are not well constrained \citep{PopescuEtAl2018_NIR-colours-spectral-type, Carry2012_DensityOfAsteroids}.
Therefore, when modelling the gravitational environment of \RS~we assume a density of $\rm 2.66~g.cm^{-3}$ - the average density of an S-type asteroid \citep{Carry2012_DensityOfAsteroids}.

We assumed a uniform density throughout the object and calculated the gravitational slope and potential of the asteroid.
The gravitational potential is displayed in Figure~\ref{fig:GravPotential} and the gravitational slopes across the surface are shown in Figure~\ref{fig:GravSlopes}.
The gravitational slopes are of most interest here, as they allow specific properties of the surface to be inferred.
Across most of the surface, the gravitational slope relative to the surface is low, with most of the surface $<20^{\circ}$, increasing to $\sim45^{\circ}$ in the neck and concave crater-like feature.
These regions could therefore suggest some level of cohesiveness or tensile strength, especially if the density is lower than estimated.
We do not observe gravitational slopes greater than $60^{\circ}$ in the current model, which would indicate exposed rock, or slopes above $90^{\circ}$, which would suggest that loose regolith would be ejected from the surface.
Whilst higher angles appear with smaller densities, a density below $\rm 1.25~g.cm^{-3}$ is required for any `negative' gravity to appear on the surface, which would be unlikely for an Sa-type object.
The higher angles in the neck and crater, however, suggest that material could move across the surface, with dust or regolith potentially accumulating at the bottom of the neck or crater.
These inferences cannot be confirmed without a better estimate of the target's surface roughness, density or composition from further spectroscopic/polarimetry measurements.

Finally, we can apply the techniques of \citet{Scheeres2007_RotationalFissionOfCBs} to estimate \RS's rotational instability.
\RS~does not have regularly shaped lobes or lobe alignment, so we use the case for two spheres, treating each lobe as a rigid body with no cohesive forces between them other than gravity.
We find that, for the rotational period of $4.445$ h, and density of $2.66\rm~g.cm^{-3}$, \RS~is stable, with a density limit of $2.21\rm~g.cm^{-3}$ required to maintain equilibrium under gravity.

\subsection{Radar Astrometry}

At the time of the initial radar track on March 12, 2014, \RS~had a well-characterized orbit solution, having optical astrometry that spanned 10 orbits since its discovery in 2000.
It therefore had been numbered by the IAU.
This resulted in the radar observations having only a minor effect on the orbit predictions of \RS.

The initial $-403.5$ $\mu s$ delay correction to orbit s58 (used for the initial Goldstone obervations) amounted to a 0.3-sigma acquisition, considering the $3844$ $\mu s$ $3\sigma$ uncertainty of s58 at the time of the measurement.
Because the orbit prediction was so close to what was subsequently measured, the collected radar measurements did not make large changes to the orbit solution.
The data's main contribution, first reflected in updated orbit solution s72, was to reduce statistical uncertainties by 25\% at the most distant statistically deterministic Earth encounters in the years 1760 and 2335.
The estimate for the next close Earth encounter in 2056 was adjusted from a nominal 0.020166 au (min. 0.020142, max. 0.020189) to s72's 0.020165 (min. 0.020162, max. 0.020167), where the uncertainty bounds are $3\sigma$ confidence values.

We later reduced additional astrometry with respect to the SHAPE-derived center-of-mass; 5 new delay measurements reported with similar 1 $\mu s$ delay uncertainties.
When included in orbit solution update s98, the SHAPE reduction was found consistent with the initial radar measurements but did not significantly change the heliocentric orbit estimate.
All radar astrometry is available here: \url{https://ssd.jpl.nasa.gov/sb/radar.html}.

\section{Conclusions} \label{sec:Conclusions}
Radar observations suggest that 15-30\% of NEAs are contact binaries \citep{ BennerEtAl2015A4_15-percent-CB-NEO, VirkkiEtAl2022_30percent-CB} with upwards of 50\% of objects in the outer solar system appearing to be either highly elongated or contact binaries \citep{ Brunini2023_50-percent-CB-Plutinos-reason}. 
Despite this, only a small selection of contact binaries have comprehensive shape models, with a wide variety of shapes and lobe arrangements \citep{CannonEtAl2025_DP14shape}.
The shape model of \RS~is an example of this, with an unusual shape compared to previously modelled contact binaries.
\RS~has a very high mass ratio between the lobes ($\sim0.86$) and its lobes are not aligned along the longest axis of the larger lobe, similar only to 67P \citep{SierksEtAl2015_67P-Shape} and Mithra \citep{BrozovicEtAl2010_Mithra-Shape-Model}.
There is a large concavity on the outside face of the larger lobe, also resembling the shape of Mithra.

The spin state of \RS~is typical of most modelled near-Earth contact binaries, with a period of $(4.445\pm0.001)$ h and a rotational pole with coordinates $\lambda = (225\pm8)^{\circ}$ and $\beta = (-80\pm9)^{\circ}$.
Such a period makes \RS~the second fastest rotator among the modelled contact binary population, behind only Castalia. 
With our new constraint on the \RS's size, we calculate an optical albedo of $p_v = 0.16 \pm 0.06$ and a radar albedo of $0.08 < \eta_{\rm OC} < 0.16$.
Whilst few Sa-types to compare to, these values place \RS~in line with other S-complex asteroids.

With the recent discovery of Selam and Donaldjohanson by the NASA Lucy mission, there is renewed interest in how these contact binary objects form.
Evidence now suggests that contact binaries could form in debris disks and through a series of collisions with other objects \citep{WimarssonEtAl_AngleOfCBs, RaducanEtAl2025_SelamFormationFromMoonlets}; however, Selam remains the only observed contact binary in a binary system, and the only known contact binary `moonlet'.
Additionally, the wide variety of shapes and locations of contact binaries within the solar system may suggest that multiple formation processes can create distinctive bilobed shapes.
Only by modelling more contact binaries can these objects be grouped and characterised further, thereby gaining insight into possible shared evolutionary histories.

Further observations of \RS~with ground-based radar to obtain further shape information will not be possible until 2056 when \RS~will come within 0.020 au of Earth, and SNRs with existing systems will be very strong.
In 2027, however, \RS~will pass within 0.2 au of Earth, allowing for the collection of additional lightcurves to confirm and update its rotational period and pole, which is currently not well constrained due to only one good epoch of observations and limited viewing geometries.

\section*{Acknowledgements}

We thank all the technical and support staff at the observatories at La Silla for their support and the staff at Goldstone and Arecibo for their help with the radar observations.
R.E.C., A.R., and C.S. acknowledge the support from the UK Science and Technology Facilities Council.
K.B. is supported by an NSF Astronomy and Astrophysics Postdoctoral Fellowship under award AST-2303858.
Research contributions by M.W.B. (SETI), L.A.M.B. (JPL), M.B. (JPL), and J.D.G. (JPL) were carried out under contract with the National Aeronautics and Space Administration (80NM0018D0004).

\section*{Data Availability}

Lightcurves are available on \href{https://alcdef.org/}{ALCDEF}.
The Arecibo data are available on request through the archive hosted by the Texas Advanced Computing Center: \url{https://tacc.utexas.edu/research/tacc-research/arecibo-observatory/}.
Goldstone radar data is available upon request, but will be made available in the future through the PDS Asteroids/Dust Subnode: \url{https://sbn.psi.edu/pds/archive/asteroids.html}.



\bibliographystyle{mnras}
\bibliography{2000rs11_ref} 




\appendix

\section{Radar modelling fits} \label{app:model fits}

\begin{figure*}
	\includegraphics[width=0.9\textwidth]{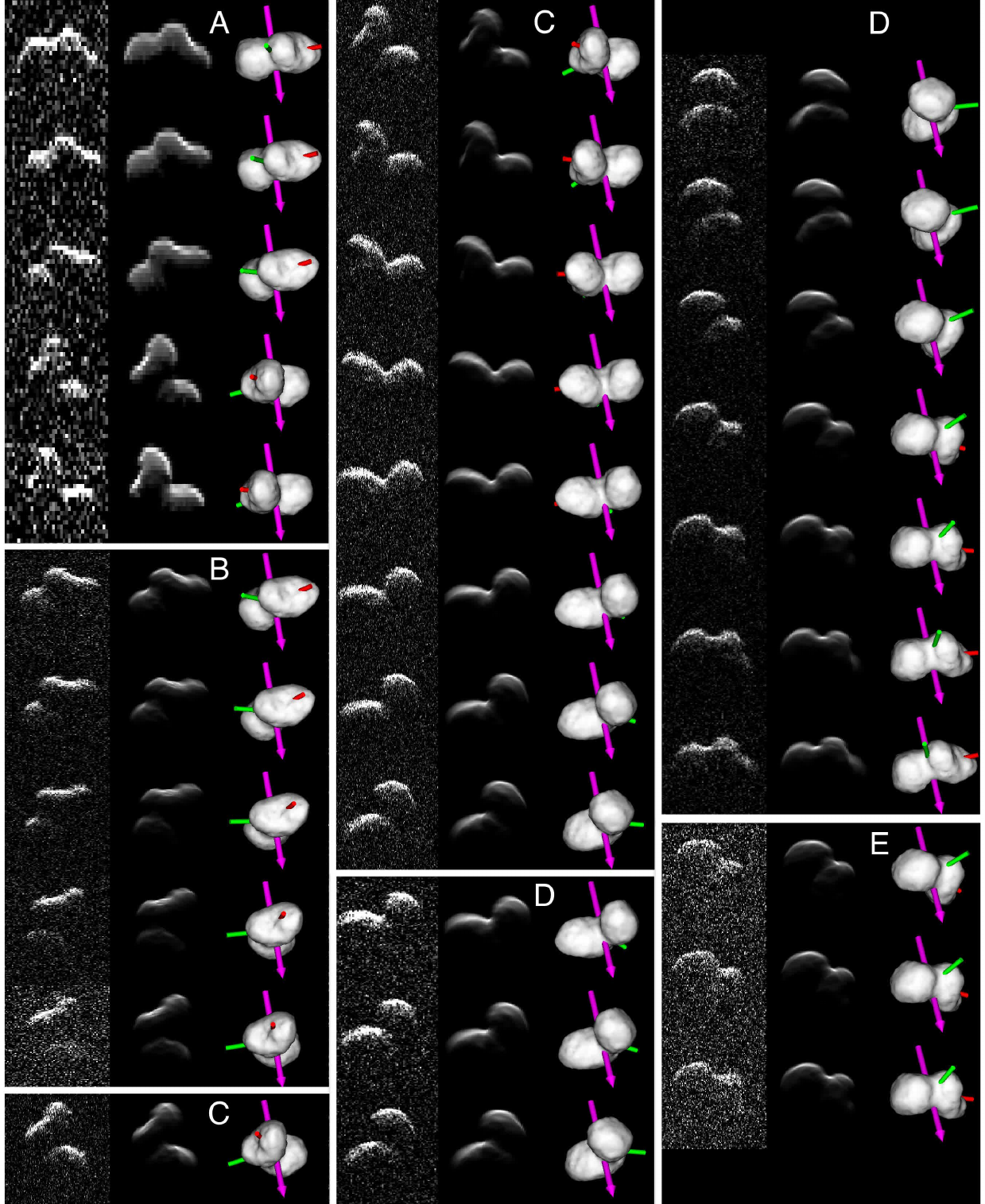}
    \caption{The raw and simulated delay-Doppler images for the final shape model. Snapshots are displayed in 3 chronological columns. Each day of data is shown in a white box as follows: A - Goldstone observations on the 13th of March, 2014; B through E - Arecibo observations on the 14th - 17th of March, 2014, respectively. Delay-Doppler images have been summed to increase SNR,introducing rotational smearing (most evident as minor blurring of the synthetic images). The three images in each column are as follows. \textit{Left}: The raw delay-Doppler images. \textit{Centre}: The simulated delay-Doppler images for the model at that time. \textit{Right}: The 900-m wide plane-of-sky view of the model, as it would be seen by direct optical imaging. The model is rotating around the pink arrow, whilst the red and green axes denote the principal axes of the model.}
    \label{figapp:radar_dd_fit}
\end{figure*}

\begin{figure*}
	\includegraphics[width=0.9\textwidth]{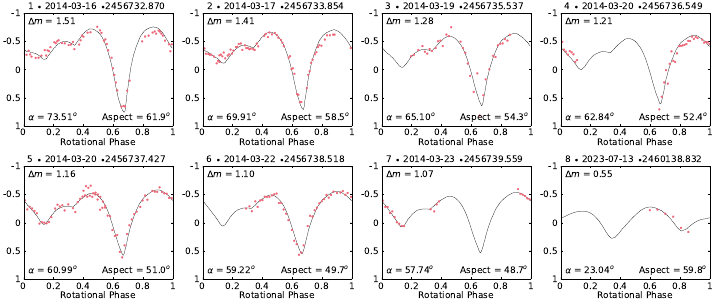}
    \caption{The lightcurve fits for the final shape model. The line shows the simulated light curve, while the dots show the observed data. For each simulated lightcurve, the peak-to-peak magnitude, solar phase angle, $\alpha$, and the aspect angle are displayed.}
    \label{figapp:radar_lc_fit}
\end{figure*}

\begin{figure*}
	\includegraphics[width=0.9\textwidth]{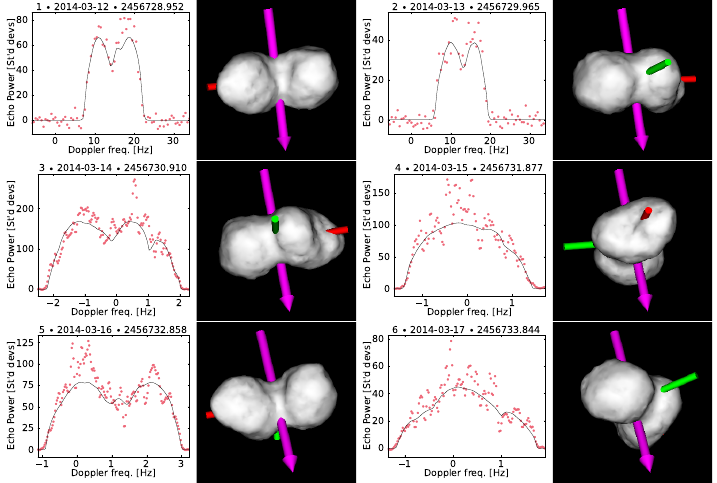}
    \caption{The OC CW radar observations of \RS~(dots) and the simulated data from the final shape model (line). 
    For each CW, the plane-of-sky view of \RS~is displayed adjacent to the right. 
    The model is rotating around the pink arrow, whilst the red and green axes denote the principal axes of the model.
    Goldstone data is displayed in X-band, and Arecibo data is displayed in S-band. 
    We did not adjust for differences in transmitter frequencies.}
    \label{figapp:radar_cw_fit}
\end{figure*}

\section{2014 Model} \label{app:2014 model}

\begin{figure*}
	\includegraphics[width=0.9\textwidth]{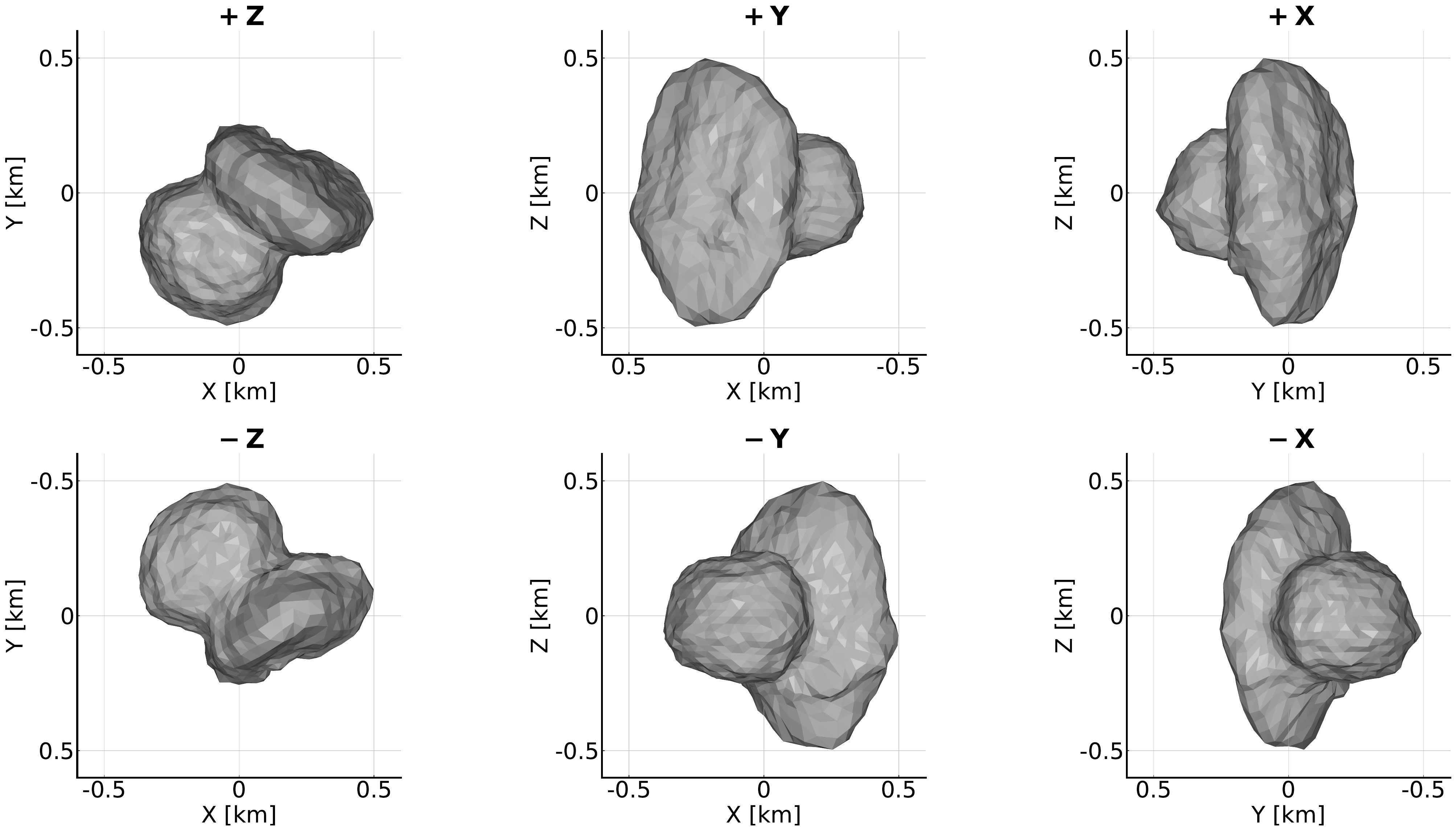}
    \caption{The 2014 model for \RS~\citep{BrauerEtAl2014_ConfAbs2000rs11_shape}, derived for the northern hemisphere solution, although delay-Doppler images alone were unable to distinguish this model between the northern and southern hemisphere solutions.}
    \label{figapp:2014_model}
\end{figure*}

\begin{figure*}
	\includegraphics[width=0.9\textwidth]{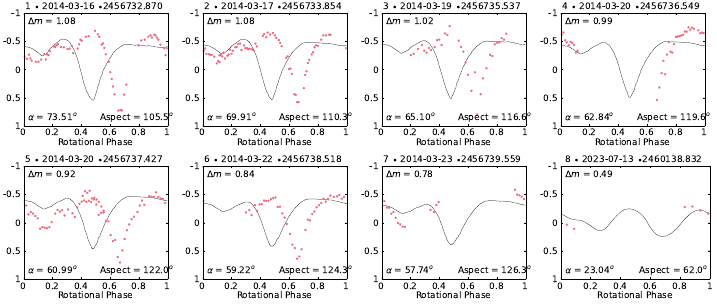}
    \caption{The lightcurve fits for the northern hemisphere solution for the 2014 shape model. We were unable to find a model that was in phase with both the radar and optical lightcurves at the same time. This figure shows the spin state that best matched the radar data, but was unable to be in phase with optical lightcurves. The line shows the simulated light curve, while the dots show the observed data. For each simulated lightcurve, the peak-to-peak magnitude, solar phase angle, $\alpha$, and the aspect angle are displayed.}
    \label{figapp:2014np_lc_fit}
\end{figure*}

\begin{figure*}
	\includegraphics[width=0.9\textwidth]{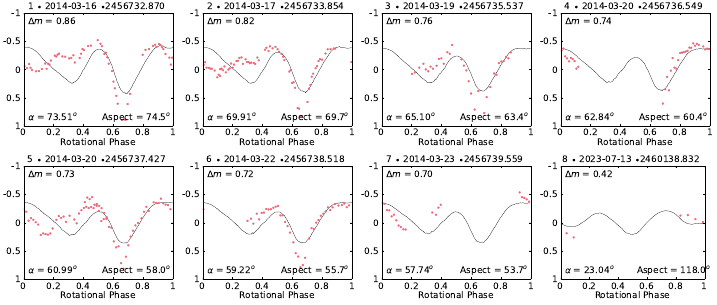}
    \caption{The lightcurve fits for the southern hemisphere solution for the 2014 shape model. The model was aligned to be in phase with the radar observations, and then independently compared to the optical lightcurves. The line shows the simulated light curve, while the dots show the observed data. For each simulated lightcurve, the peak-to-peak magnitude, solar phase angle, $\alpha$, and the aspect angle are displayed.}
    \label{figapp:2014sp_lc_fit}
\end{figure*}


\bsp	
\label{lastpage}
\end{document}